\documentclass[manuscript]{aastex}

\newcommand{\angeo}{AnGeo}
\newcommand{\soph}{SoPh}
\newcommand{\ssrv}{SSRv}
\newcommand{\jgra}{JGR}
\newcommand{\grla}{GRL}
\newcommand{\aaa}{A\&A}
\newcommand{\prla}{PhRvL}

\slugcomment{Not to appear in Nonlearned J., 45.}

\shorttitle{Strahl widths in presence of whistler waves}
\shortauthors{Kajdi\v{c} et al.}

\begin{document}


\title{Suprathermal electron strahl widths in the presence of narrow-band whistler waves in the solar wind}

\author{P. Kajdi\v{c}\altaffilmark{1}}
\affil{Instituto de Geof\' isica, Universidad Nacional Aut\'onoma de M\'exico, Mexico City, Mexico}
\email{primoz@geofisica.unam.mx}

\author{O. Alexandrova\altaffilmark{2}, M. Maksimovic\altaffilmark{2} and C. Lacombe\altaffilmark{2}}
\affil{LESIA, Observatoire de Paris, PSL Research University, CNRS, UPMC Universit\'e ́ Paris 06, Universit\'e ́ Paris-Diderot, 5 Place Jules Janssen, F-92190 Meudon, France}

\author{A. N. Fazakerley\altaffilmark{3}}
\affil{Mullard Space Science Laboratory, University College London, UK}

\begin{abstract}
We perform the first statistical study of the effects of the interaction of suprathermal electrons with narrow-band whistler mode waves in the solar wind.  We show that this interaction does occur and that it is associated with enhanced widths of the so called strahl component. The latter is directed along the interplanetary magnetic field away from the Sun. We do the study by comparing the strahl pitch angle widths in the solar wind at 1AU in the absence of large scale discontinuities and transient structures, such as interplanetary shocks, interplanetary coronal mass ejections, stream interaction regions, etc. during times when the whistler mode waves were present and when they were absent. This is done by using the data from two Cluster instruments: STAFF data in frequency range between $\sim$0.1~Hz and $\sim$200~Hz were used for determining the wave properties and PEACE datasets at twelve central energies between $\sim$57~eV (equivalent to $\sim$10 typical electron thermal energies in the solar wind, E$_T$) and $\sim$676~eV ($\sim$113~E$_T$) for pitch angle measurements.
Statistical analysis shows that during the intervals with the whistler waves the strahl component on average exhibits pitch angle widths between 2$^\circ$ and 12$^\circ$ larger than during the intervals when these waves are not present. The largest difference is obtained for the electron central energy of $\sim$344~eV ($\sim$57~E$_T$).

\end{abstract}


\section{Introduction}

It was discovered very early that the electron velocity distribution function (VDF) in the solar wind is composed of different components \citep[e.g.][]{montgomery68, feldman75, feldman78, rosenbauer76, lin98, maksimovic05}.
About 95~$\%$ of all the electrons belong to the thermal core population with typical temperature of $\sim$10~eV. These electrons are marginally collisional at 1~AU \citep{phillipsgosling90} and their VDF can be described as bi-Maxwellian (one in parallel and one in perpendicular directions with respect to the mean field).

Electrons with energies between $\sim$70~eV and $\sim$2~keV are referred to as suprathermal. These are collisionless at 1~AU \citep{scudderolbert79, fairfieldscudder85,ogilvie00}, so they are not in thermodynamic equilibrium. Suprathermal electrons are composed of two components: halo exhibits approximately isotropic VDF with suprathermal tails that can be approximated by a bi-kappa distribution \citep[e.g.][]{maksimovic97, maksimovic05, stverak09}. The strahl component can be described as field aligned anti-sunward directed beam of electrons.

Finally, the electrons with energies above 2~keV compose isotropic superhalo \citep{lin98}.

There has been some discussion about the origin of suprathermal electrons. \citet{pierrard99}  used the {\it Wind} observations of the electron VDF at 1~AU to derive the coronal VDF and concluded that suprathermal electrons must already be present in the corona.
\citet{vocksmann03}, \citet{vocks08} and \citet{vocks12} postulate that the suprathermal population is formed in the inner corona by resonant interaction with antisunward propagating whistler waves. These waves scatter the sunward propagating portion of core electrons from small velocities parallel to magnetic field ($v_{\parallel}$) to large perpendicular velocities ($v_{\perp}$) thereby forming the halo.

Whatever the origin, the antisunward propagating suprathermal electrons (in the plasma frame) are subject to focusing effects by the diverging interplanetary magnetic field (IMF) due to conservation of the particle's magnetic moment \citep[e.g.,][]{owens08}. If no other effects were present in the interplanetary (IP) space, these electrons would be focused into a very narrow beam or strahl. Observations however show strahl with a finite width \citep[e.g. ][]{fitzenreiter98}. Hence, some mechanism(s) must exist in the IP space which scatter the strahl electrons towards larger pitch-angles (PA). 

It is commonly postulated that halo at large heliocentric distances is formed by scattering of the strahl electrons. Some indirect evidence point in this direction: \citet{maksimovic05} and \citet{stverak09} for example have shown that while the core fractional density remains constant with the distance from the Sun, the halo and the strahl fractional densities vary in opposite ways. The halo fractional density increases with increasing heliocentric distance, that of the strahl decreases, while their sum remains roughly constant. 

Electromagnetic fluctuations (frequency $\omega$) can resonantly interact with electrons in the solar wind if their Doppler shifted frequency in the electron frame is equal to a multiple of the electron cyclotron frequency $\Omega_e$. This resonance condition reads:
\begin{eqnarray}
\omega - k_{\parallel}v_{\parallel} = n\Omega_e; n = 1, 2, 3, ...,
\end{eqnarray}
\noindent where $k_{\parallel}$ and $v_{\parallel}$ are the components of the wave vector and electron velocity parallel to the background magnetic field. 

Whistler waves, which have frequencies $\omega < \Omega_e$ and a right-handed polarization with respect to the background magnetic field \citep[e.g.][]{gurnett05}, can resonate with electrons if $k_{\parallel}\cdot v_{\parallel}$ is negative: antisunward propagating electrons can only interact with sunward propagating whistler waves.

Two potential sources of whistler mode fluctuations in the solar wind are \citep[e.g., ][]{saitogary07} wave-particle interactions and wave-wave interactions. The first can generate whistler fluctuations through electromagnetic instabilities such as heat flux instability and the electron temperature anisotropy instability. The wave-wave interactions may result in magnetic fluctuations cascading. It is well known
that at low frequencies the magnetic power spectrum in the solar wind exhibits frequency dependence $f^{-5/3}$, \citep[e.g. ][]{bruno13}. At around the proton cyclotron frequency the spectrum becomes steeper  \citep[e.g. see the review of][]{alexandrova13}. The nature of this small scale turbulent cascade is still an open question. Some authors, e.g. \citep{denskat83, ghosh96, stawicki01, smith06a}, suggest that 
fluctuations in this range may be whistler mode waves with broad spectrum (as opposed to narrow-band whistler wave modes described here in this paper). 

Broadband whistler waves propagating parallel to the background B-field were introduced in simulations by \citet{vocks05} who showed that in IP space these waves could indeed disperse the strahl.
\citet{pierrard11} also proposed that the strahl electrons could be scattered off the whistler broadband turbulence with wave vectors parallel to the background magnetic field. However, observations show that within this small scale range, turbulent fluctuations are dominated by quasi-perpendicular wave vectors $k_{\perp} \gg k_{\|}$ \citep{mangeney06,alexandrova08,chen10,sahraoui10,roberts13} and not by quasi-parallel ones as needed in the previously mentioned models (see discussion in section~4 for more details). 
 Alternatively, \citet{pavan13} suggested that self-generated Langmuir waves at plasma frequency could also scatter the strahl in picth angle and energy, resulting in significant broadening of its VDF.

Direct observations of halo formation from the strahl component have been reported by \citet{gurgiolo12}. These authors exmined electron velocity distribution functions obtained by the PEACE instrument onboard the Cluster spacecraft.
\citet{gurgiolo12} show a handful of time intervals during which scattering of the strahl into what they call the proto-halo and then into the halo was observed. This occured for electrons at energies $\lesssim$50~eV during time intervals of $\sim$10 seconds. The authors also examined magnetic field turbulence from the STAFF and FGM datasets and concluded that no monochromatic whistler mode waves were present during the examined intervals but that there were enhanced levels of broadband turbulence.

In contrast to previous works, we study the broadening of the strahl during times when narrow-band whistler waves are present in the solar wind. By narrow-band we mean that in the spectra of magnetic field turbulence, these waves produce a clear, distinct bump, which is superimposed on the spectra of permanent background turbulence.
Recently \citet{lacombe14} performed a study of such waves and determined preferential conditions in the solar wind for their observations. These include a low level of background turbulence, a slow wind, a relatively large electron heat flux, and a low electron collision frequency. The authors related the presence of the whistlers preferentially to the whistler heat flux instability and in rare cases to the anisotropy instability of the total electron temperature.

This paper is organized in the following manner: In section 2 we describe the instruments and the datasets used in this work and present a case study. In section 3 we discuss the properties of the IMF and the solar wind during times intervals in our sample, the observational properties of the whistler waves and the measured strahl widths. In section 4 we discuss the results and summarize them.

\section{Observations}
\subsection{Instruments and datasets}
The Cluster mission consists of four identical spacecraft in orbit around the Earth. It provides magnetic field and plasma measurements in the near-Earth environment. The satellites carry several instruments onboard. Here we use the data provided by five instruments: the Fluxgate Magnetometer \citep[FGM,][]{balogh01}, the Cluster Ion Spectrometer \citep[CIS,][]{reme01}, the Plasma Electron And Current Experiment, \citep[PEACE,][]{johnstone97}, the Spatio Temporal Analysis of Field Fluctuations experiment \citep[STAFF,][]{cornilleauwehrlin97, cornilleauwehrlin03} and the  Waves of High frequency and Sounder for Probing of Electron density by Relaxation \citep[WHISPER,][]{decreau97}.

All the data used in this work were obtained from the Cluster Science Archive \\(http://www.cosmos.esa.int/web/csa), which is maintained by the European Space Agency.
We use the FGM magnetic field vectors and the CIS-HIA solar wind ion moments with 0.2~second and 4~second time resolution, respectively. To obtain electron pitch angle distributions we use
PEACE PITCH\_SPIN datasets. {\bf These contain data from both PEACE sensors, namely the High Energy Electron Analyser (HEEA) and the Low Energy Electron Analyser (LEEA). The data in them are binned in twelve 15$^\circ$ pitch angle bins and 44 energy bins.} In our work we use approximate central energies of 676~eV, 536~eV, 430~eV, 344~eV, 276~eV, 220~eV, 175~eV, 140~eV, 111~eV, 89~eV, 71~eV and 57~eV. {\bf We note here that the energies in PITCH\_SPIN datasets are not corrected for spacecraft potential, however during our intervals the potential was typicaly less than 5~V, which is far less than the energy intervals used here. The central energies between 57~eV and 676~eV were chosen since we find that at higher energies the PADs become too noisy and not many good examples could be obtained. The lower threshold was chosen since the usual breakpoint between the core and the suprathermal electrons is around 60~eV \citep{feldman75}.} 
PEACE data are available in 4~second time resolution. We use WHISPER data in order to make sure that the Cluster probes are not located inside the Earth's foreshock. This is done by checking for the presence of the electrostatic or Langmuir waves which are commonly present in the foreshock.

The STAFF experiment measures the three orthogonal components of the magnetic field fluctuations. It comprises two onboard analyzers: the wave form unit (STAFF-SC) provides digitized wave forms up to either 12.5~Hz or 180~Hz, depending on the spacecraft telemetry rate. The Spectrum Analyzer (STAFF-SA) uses the three magnetic field and two electric field components \citep[from the EFW experiment,][]{gustafsson97} to build a 5$\times$5 spectral matrix, which in normal telemetry rate has a time resolution of 4 s and the frequency range between 8~Hz and 4~kHz. The two analyzers provide the sense of polarization, the ellipticity and the propagation direction of the observed fluctuations. At times, when the measurements of electric field are of good quality, it is possible to determine the sense of the wave vector without the 180$^\circ$ ambiguity.

{\bf The data are from C1, C2 and C4 spacecraft with the waves, the electrons and the magnetic fields measured by the same spacecraft. However the ion moments from C2 and C4 are not available for time intervals in our sample (Table~\ref{tab1}). In these time cases we first compare the B-field data of C1 and C2 to see whether the two spacecraft were close enough to each other in order to observe the same regions in space. If this is true then we use C1 ion moments for calculating plasma parameters, such as electron gyrofrequency, etc. Table~\ref{tab1} shows information on the spacecraft that provided the data for each time interval.}

\subsection{Case study}
In this section we describe one case study in order to explain our methodology.
We surveyed the data of the Cluster mission during the years 2001-2010. We searched for times when the Cluster was in the pristine solar wind (SW), meaning that the satellites were not in the Earth's foreshock nor was the SW perturbed by transient structures, such as stream interaction regions or interplanetary coronal mass ejections. 

\begin{figure*}
\begin{center}
\includegraphics[width = 0.6\textwidth]{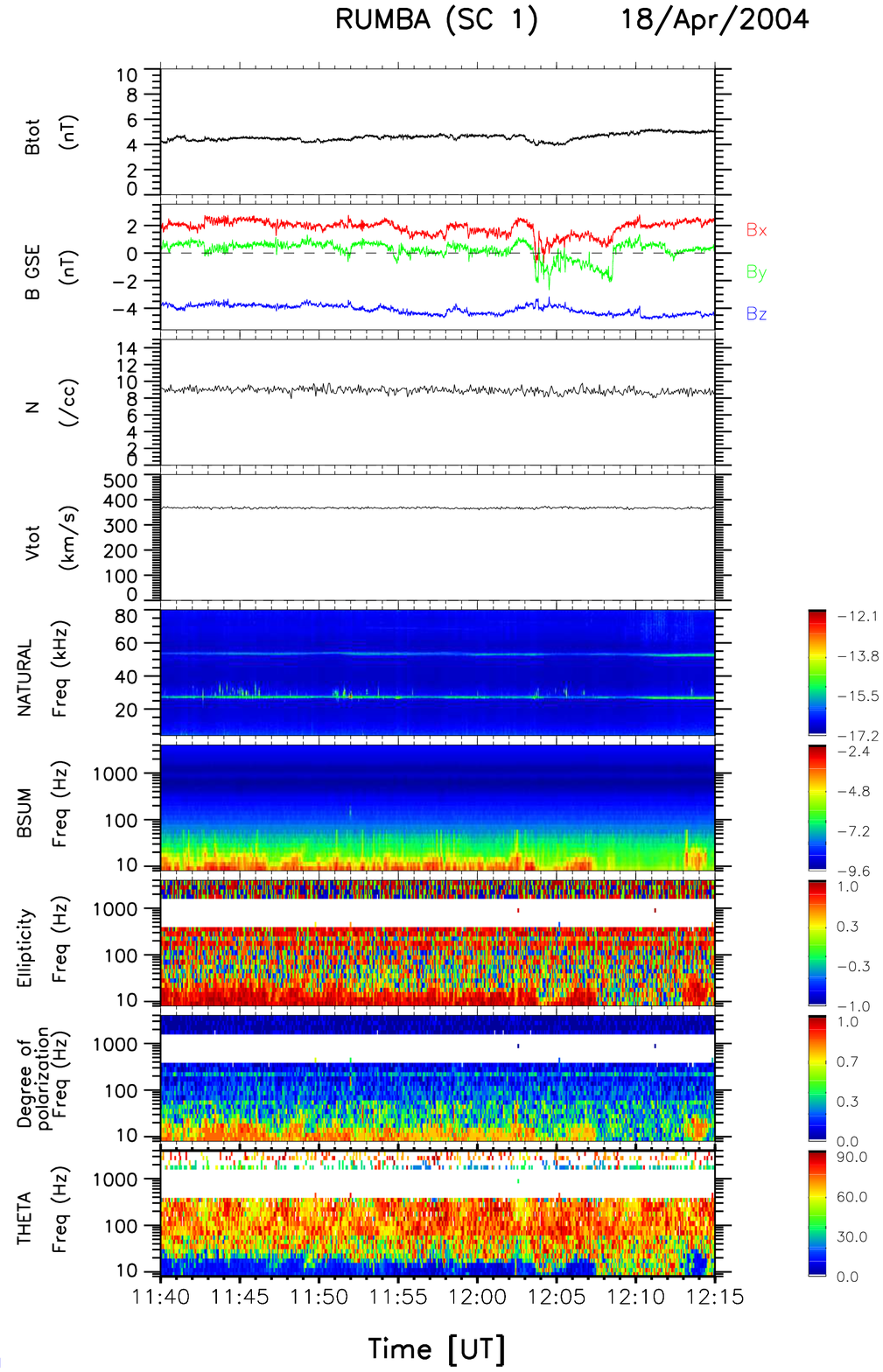}
\caption{An example of whistler waves on 18 April, 2004 observed by the Cluster 1 spacecraft. The panels show (from top to bottom): magnetic field magnitude and components in GSE coordinates (in units of nT), the solar wind number density (cm$^{-3}$) and the solar wind velocity (kms$^{-1}$). The fourth panel exhibits the  electric field dynamic spectrum from the WHISPER. The lower four panels exhibit (from top to bottom): the dynamic spectrum of total energy of magnetic field fluctuations B$_{SUM}$, ellipticity (+1 for right-hand and -1 for left-hand polarized fluctuations), degree of polarization (0 = linear, 1 = circular) and the angle of propagation (between the wave vector $k$ and the IMF, $\theta_{kB}$).}
\label{fig:sample}
\end{center}
\end{figure*}

We use STAFF datasets in order to search for elliptically polarized, right-hand fluctuations, that propagate at small angles with respect to the background IMF. Figure~\ref{fig:sample} shows Cluster 1 observations from 11:40~UT to 12:15~UT on 18 April 2004. 
The top four panels show the interplanetary magnetic field magnitude and GSE components in units of nanoTesla (nT), the solar wind number density (cm$^{-3}$) and the solar wind velocity (kms$^{-1}$). The fourth panel shows the electric field dynamic spectrum from WHISPER. The lower four panels show the STAFF data: the dynamic spectrum of total energy of magnetic field fluctuations B$_{SUM}$, ellipticity (+1 for right-hand and -1 for left-hand polarized fluctuations), degree of polarization (0 = linear, 1 = circular) and the angle of propagation (between the wave vector $k$ and the IMF, $\theta_{kB}$). 
During the presented time interval the spacecraft is in the pristine solar wind. The IMF and plasma properties are stable throughout the interval. The STAFF data show continuous B-field fluctuations in frequency range between 8~Hz and $\sim$20~Hz throughout the shown time intervals except between 12:07:30~UT and 12:13:00~UT. These fluctuations are right-handed, they are elliptically polarized and propagate with an angle $\theta_{kB} <$ 30$^\circ$, hence we classify them as whistler mode waves.

\begin{figure*}
\begin{center}
\includegraphics[width = 0.45\textwidth]{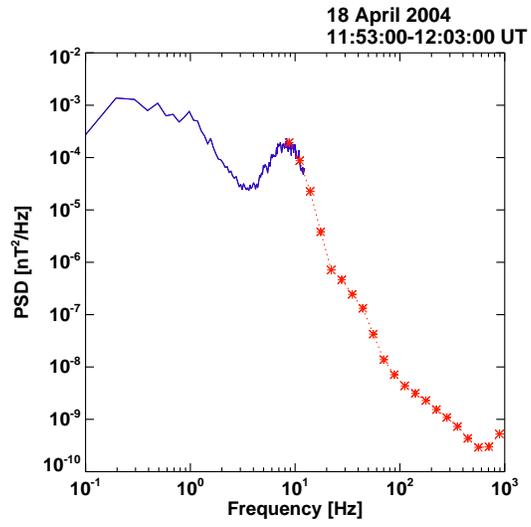}
\caption{Whistler wave spectra on 18 April, 2004. The STAFF-SC data are represented with a purple continuous line, while red asterisks are for the STAFF-SA data.}
\label{fig:sampleA}
\end{center}
\end{figure*}

Figure~\ref{fig:sampleA} shows the average spectrum of magnetic fluctuations during the time interval from 11:53:00~UT to 12:03:00~UT. This spectrum is obtained by calculating the average power of fluctuations at each central frequency during that time. We plot the STAFF-SC part of the spectrum (below $\sim$12.5~Hz) with continuous purple line, while the asterisks represent the STAFF-SA data (above $\sim$12.5~Hz). The spectrum consists of continuous part, which belongs to background turbulence and a distinct bump centered at around 8~Hz, which is due to the whistler mode waves.

\begin{figure*}
\begin{tabular}{cc}
\includegraphics[width = 0.55\textwidth]{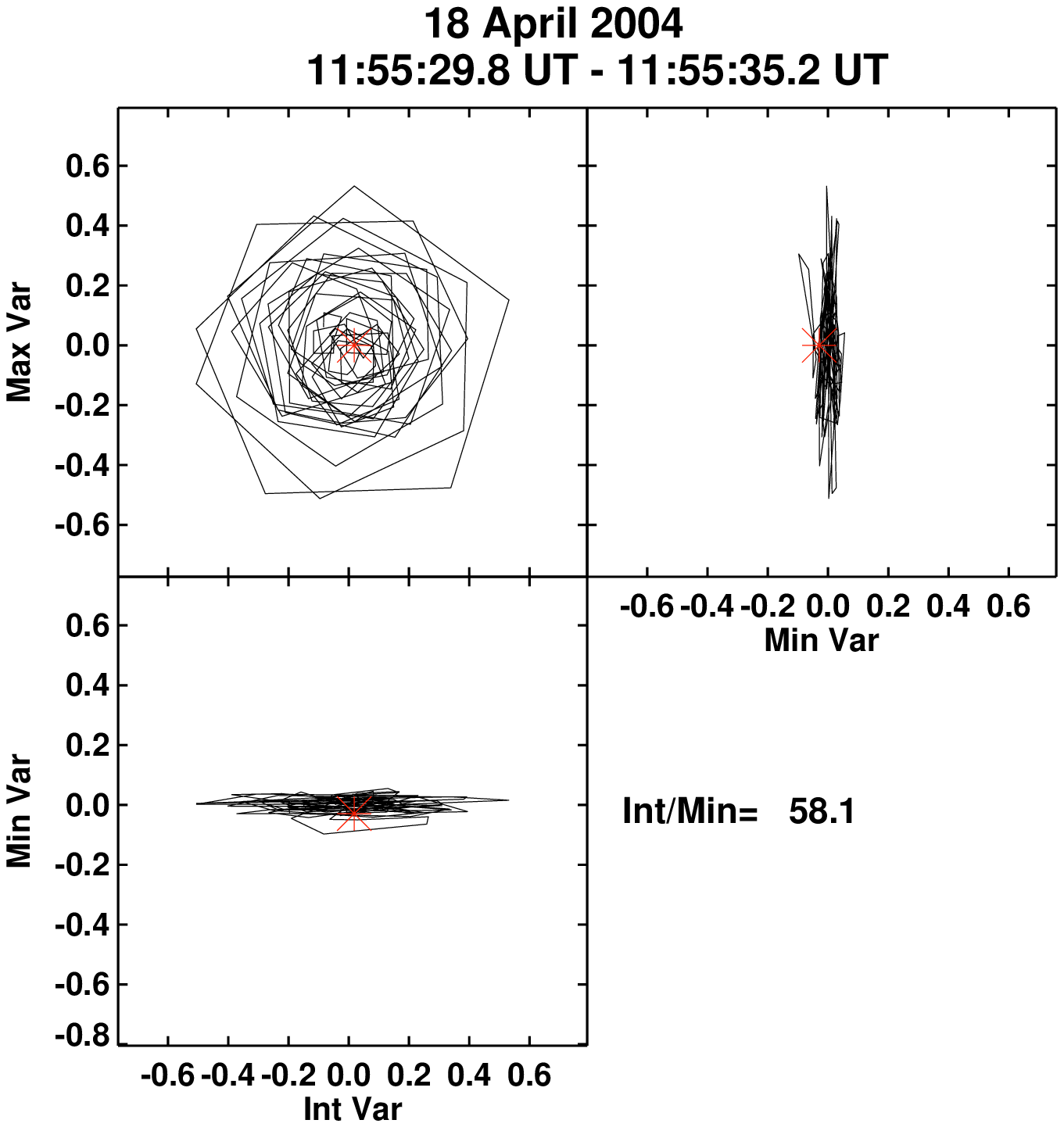} & \includegraphics[width = 0.55\textwidth]{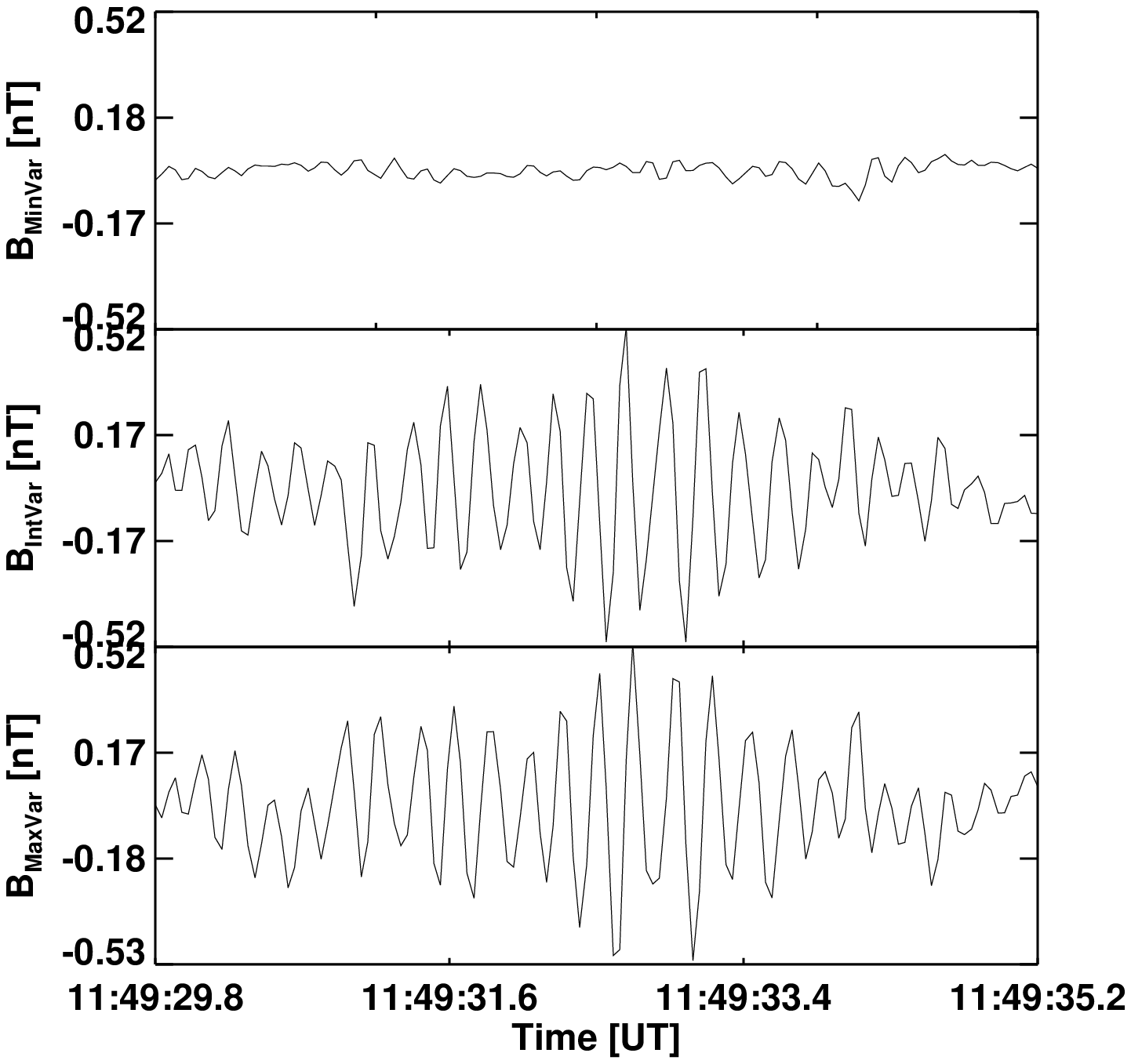}\\
{\bf a)} & { b)}
\end{tabular}
\caption{(a) Hodograms of whistler waves observed on 18 April, 2004 between 11:55:28.9-11:55:35.2~UT. It can be seen that these are very planar waves with the intermediate to minimum variance ratio of 58. The red asterisk marks the begining of the time interval. (b) Whistler waveforms in the coordinate system defined by the eigenvectors from the minimum variance analysis. The panels show (from top to bottom): B-field components in the minimum, the intermediate and the maximum variance directions.}
\label{fig:sample1}
\end{figure*}

Figure~\ref{fig:sample1}a shows the results of the minimum variance analysis \citep[MVA, ][]{sonnerup98} performed on the STAFF-SC data with 25~s$^{-1}$ time resolution between 11:55:29.8~UT and 11:55:35.2~UT. It can be seen that the whistler waves are highly planar with the ratio of the intermediate and the minimum (Int/Min) variances of 58. In Figure~\ref{fig:sample1}b we show waveforms of these waves in the frame of eigenvectors obtained from the MVA. The panels (from top to bottom) show B-field profiles in the direction of the minimum, the medium and the maximum variance.

Next, we use the PEACE PITCH\_SPIN\_DEFlux datasets in order to perform the measurements of the strahl width. These datasets contain the electron differential energy flux (DEF) as a function of the pitch angle (between the particle's velocity vector and the IMF) at spin (4-second) time resolution for different central energies. The DEF is a product of the differential particle flux (DF) times the particle energy. The DF measures the number of particles with energy $dE$ about $E$ with direction $d\vec{\Omega}$ about $\vec{\Omega}$ that passes through the unit area perpendicular to $\vec{\Omega}$ per unit time. Its units are $1/(cm^2\cdot s\cdot str\cdot eV)$, while the DEF is measured in units of $eV/(cm^2\cdot s\cdot str\cdot eV$).

We calculate the average pitch angle distribution (DEF vs. PA) during one minute time intervals at twelve central energies between $\sim$57~eV and $\sim$676~eV. We fit the PA distributions  with a Gaussian function described by equation~\ref{eqn:02}: 

\begin{eqnarray}
f(PA, w) = f_{halo} + f_{0, strahl}e^{-\big(\frac{PA-PA_0}{\sqrt{2}w}\big)^2}.
\label{eqn:02}
\end{eqnarray}
\noindent Here, $f_{halo}$ represents the constant contribution of the halo component, while the second term approximates the strahl distribution. PA stands for pitch angle (angle between the particle's velocity vector and the background magnetic field) and $w$ represents the width of the strahl centered at PA$_0$, which can have values of 0$^\circ$ or 180$^\circ$.
The fitting was performed by using the IDL CURVEFIT function. This function uses a gradient-expansion algorithm in order to provide a non-linear least squares fit to any function with arbitrary number of arguments. We adapted the CURVEFIT function in order to obtain the best estimates of $f_{halo}, f_{0,strahl}$ and $w$ and also of their errors.

\begin{figure*}
\begin{tabular}{cc}
\includegraphics[width = 0.45\textwidth]{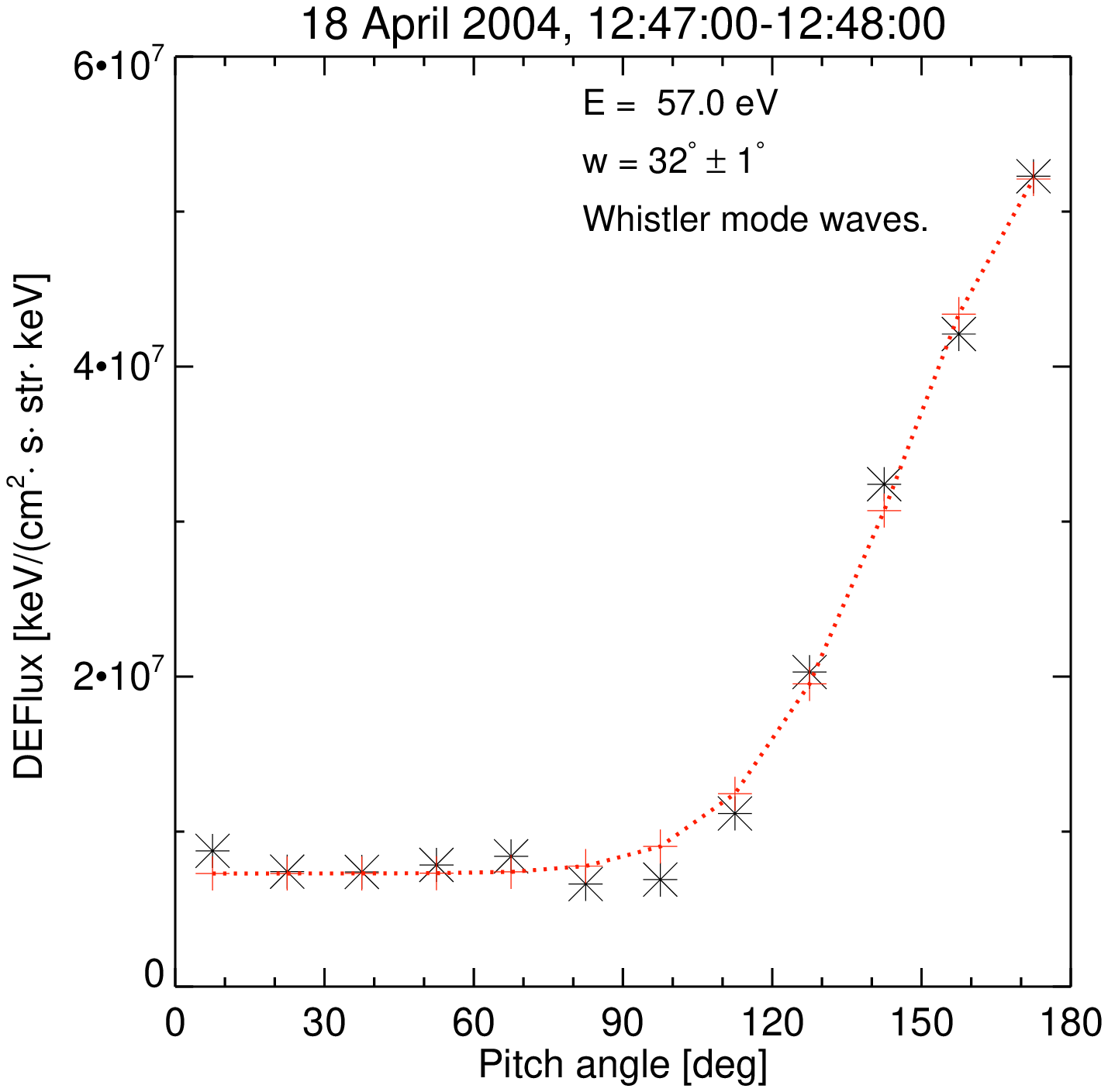} & \includegraphics[width = 0.45\textwidth]{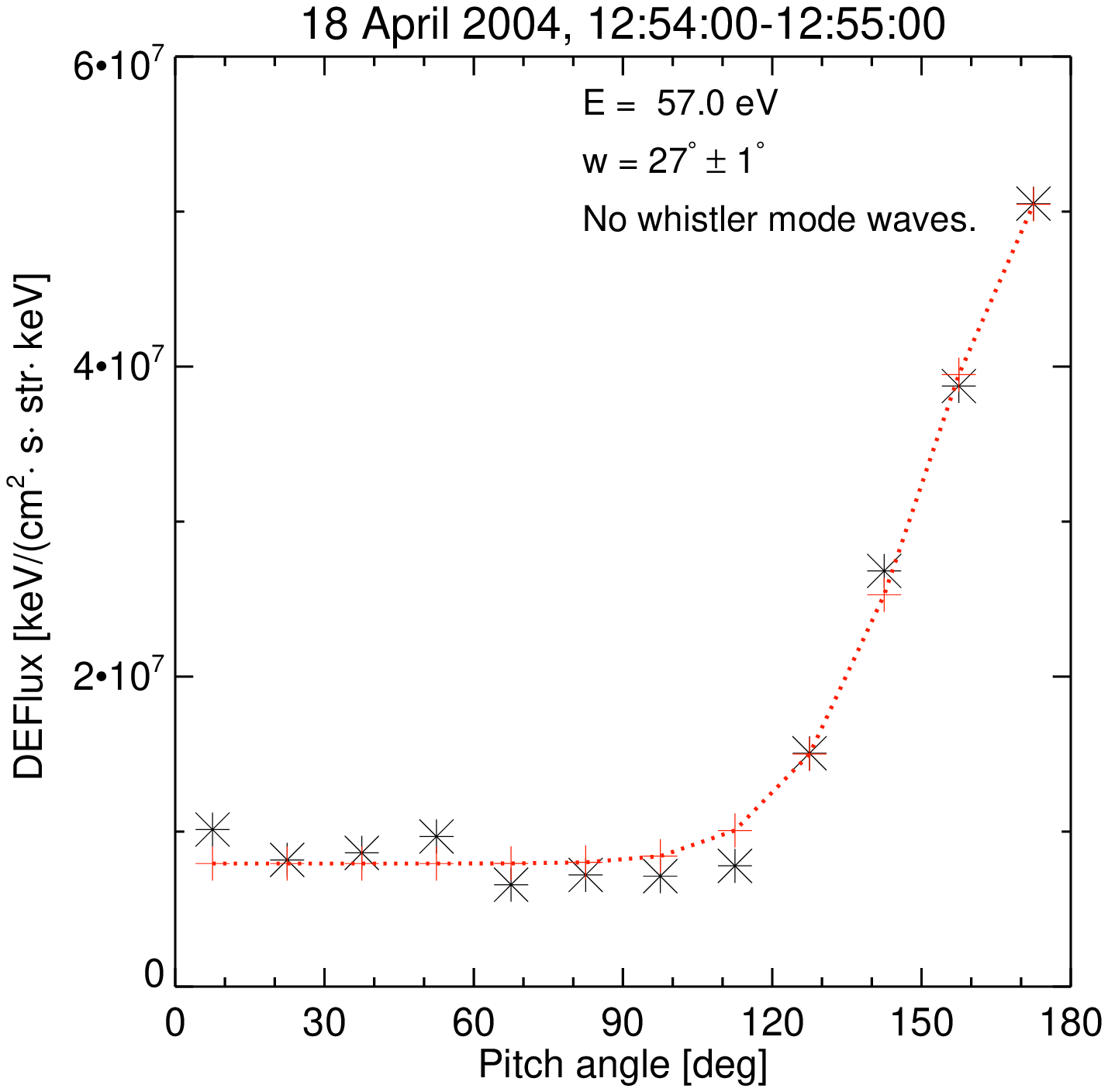} \\
a) & b)
\end{tabular}
\caption{Examples of PA distributions and the corresponding fits on 18 April, 2004 during one minute time intervals when the whistler mode waves were present (left) and when they were absent (right). Black asterisks show PADs from the data, while red crosses and dotted lines show the best fits.}
\label{fig:pad}
\end{figure*}

Figure~\ref{fig:pad} shows two examples of PADs observed on 18 April, 2004, which are separated by a few minutes. On both panels the black asterisks represent the PADs from the data, while red crosses and dotted lines represent the best fits. The time intervals, the central energies and the widths of the fits are shown on the panels. We can see that during the time interval when the whistler mode waves were present (left), the fitted width of the PAD is $\sim$32.2$^\circ$ while during the time when the whistlers were absent, it is $\sim$27.4$^\circ$. This is a large difference and it is much larger than the estimated width errors, {\bf which are $\sim$1$^\circ$. The latter value is typical for our set of PADs.}

{\bf We should state here that we visually inspected all the PADs in our sample in order make sure that the halo and the strahl components were present. This is important since most of our samples were observed during the slow solar wind (V$_{sw}\lesssim$400~kms$^{-1}$), while the strahl component has been recognized to be permanent feature only in high-speed solar wind stream \citep{rosenbauer77, feldman78}. However, \citet{pilipp87b} showed that the strahl can be observed also in the slow solar wind.}

\section{Statistical results}

\subsection{IMF and solar wind}

Here we briefly discuss the properties of the solar wind and the interplanetary magnetic field during time intervals in our samples. 
All the intervals were selected so that the Cluster spacecraft were located in the pristine solar wind far from any discontinuities, such as interplanetary shocks, they were not inside the Earth's foreshock nor within any transient structures, such as interplanetary coronal mass ejections (ICME) and stream interaction regions (SIR). Transient structures, such as SIRs are associated with enhanced magnetic field magnitudes. As suprathermal strahl electrons propagate into such regions, their PA distributions become wider due to conservation of the electron magnetic moment and this could interfere with our study.

Another reason for avoiding transient structures and IP shocks is that due to enhanced B-field magnitudes associated to them, some of the strahl electrons may be subject to adiabatic mirroring and propagate sunward at some acute PAs (between 0$^\circ$ and 90$^\circ$). Such populations of suprathermal elecrons are known as conics \citep[e.g., ][]{gosling01}. As these electrons approach the Sun, they are again reflected due to strong magnetic field there and form another population called shoulders \citep{gosling01} at PAs that are complementary to those of the conics. Associated to conics and shoulders are also halo depletions which are centered at 90$^\circ$ PA \citep{gosling01, lavraud10}. Suprathermal conics and halo depletions were also observed inside ICMEs \citep{feldman99, gosling02}. Additionally, \citet{kajdic14} reported observations of 90$^\circ$ PA enhancements near many IP shocks. We avoid all these electron signatures and select time intervals without them.

\begin{figure}
\centering
\includegraphics[width = 0.9\textwidth]{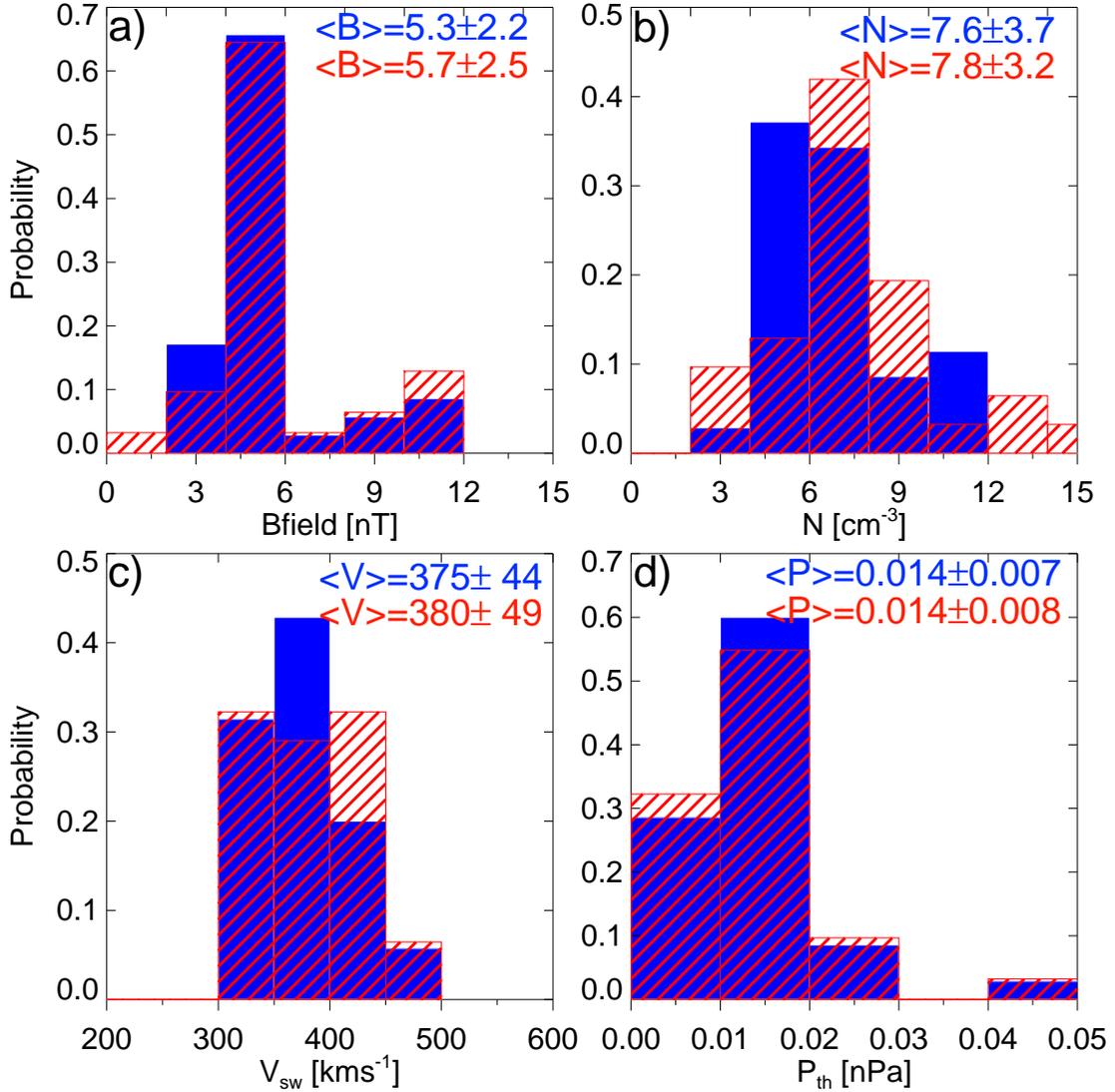}
\caption{IMF and solar wind properties during time intervals when the whistler mode waves were present (blue columns) and when they were absent (red, hashed columns). The following quantities are shown: a) IMF magnitude, b) solar wind density, c) solar wind velocity and d) solar wind thermal pressure. The averages and standard deviations are also shown in the Figure.} 
\label{fig:compare}
\end{figure}

Average observational properties of SW and IMF from our sample were very similar regardles of whether whistler mode waves were present or not {\bf (see Figure~\ref{fig:compare})}. The observed IMF magnitudes ranged between 1~nT and 12~nT with the most commom value at around 5~nT. All whistler waves were observed during times of slow solar wind (v$\leq$500~kms$^{-1}$) when the thermal pressure P$_{th}$ was $<$0.05~nPa. The plasma density displayed average value of 7.8~cm$^{-3}\pm$3.2~cm$^{-3}$. Finally, the angle between the SW bulk velocity and the IMF ($\theta_{BV}$, not shown) was always above 45$^\circ$. This has to do with the orbit of the Cluster mission. As explained by \citet{alexandrova12}, when $\theta_{BV}$ is large there is more probability that the Cluster will not be magnetically connected to the Earth's bow shock (so it will be in pristine solar wind). For the whistler waves this also means that, since they propagate at small angles with respect to the background B-field, their frequencies in the spacecraft frame of reference will not be strongly Doppler shifted.
The results on IMF and SW properties match well those reported by \citet{lacombe14} during their observations of the whistler mode waves. 

\subsection{Whistler wave properties}
\begin{figure*}
\includegraphics[width = 0.9\textwidth]{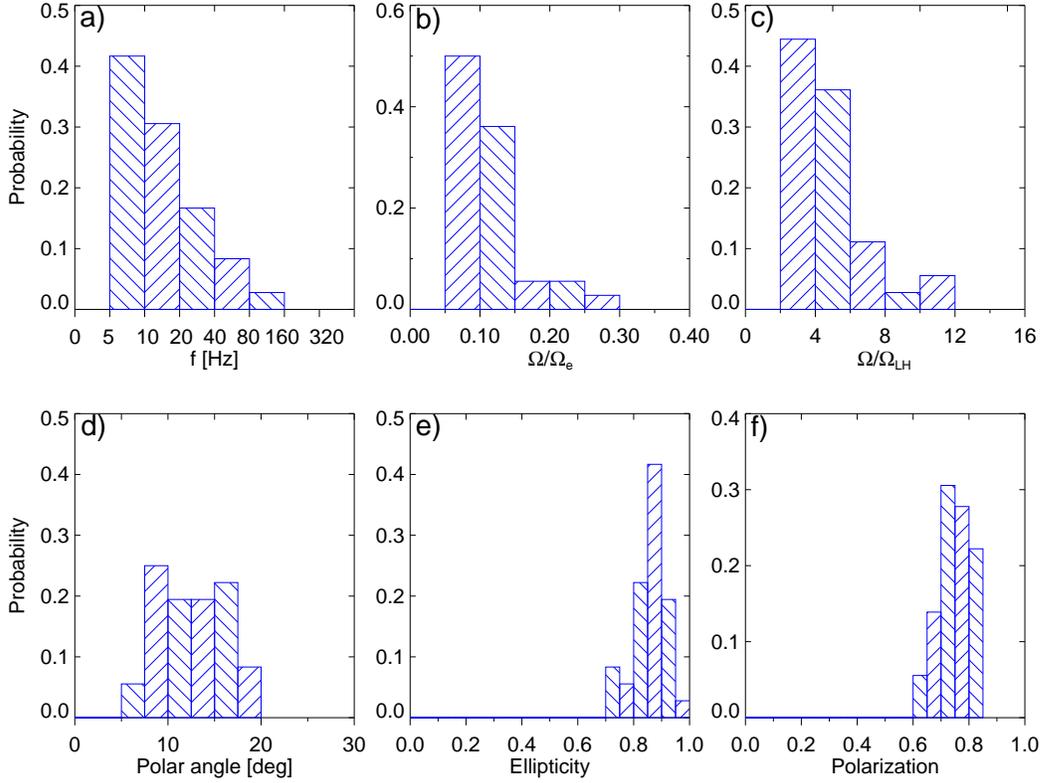}
\caption{Distributions of observed whistler wave properties: a) observed frequency, b) whistler frequency normalized to the electron gyrofrequency, c) whistler frequency normalized to lower hybrid frequency, d) angle of propagation with respect to the background IMF, $\theta_{kB}$, e) ellipticity and f) degree of polarization.}
\label{fig:whistlers}
\end{figure*}

Figure~\ref{fig:whistlers}  exhibits the distributions of the properties of the observed whistler waves. On panel a) are shown their peak frequencies. These are all $\geq$5~Hz and tend to be less than 50~Hz, although a few examples have been found at higher frequencies. We also show the whistler frequencies in units of electron gyrofrequency ($\Omega_e$, panel b) and the lower hybrid frequency ($\Omega_{LH}$, panel c). The observed $\omega/\Omega_e$ values range between $\sim$0.05 and 0.3 with most of them being below 0.2. The $\omega/\Omega_{LH}$ values are between $\sim$2 and $\sim$7.

The bottom three panels exhibit propagation properties of the observed whistler waves. The values shown are the averages of the propagation properties during the selected time intervals for central frequency at which the whistler average wave spectra peak. In panel d) we show the angle of propagation with respect to the IMF ($\theta_{kB}$), which is between 5$^\circ$ and 20$^\circ$. On panel e) we show the ellipticity, where positive values mean right-hand polarized waves and negative values mean left-hand polarization. In all of our cases the polarization is right-handed, as is should be for the whistler waves, with values between 0.7 and 1.
Finally, panel f) shows the degree of polarization  \citep[see ][for details on polarization and ellipticity of waves from STAFF-SA datasets]{santolik03}, where values close to 0 mean waves that are linearly plarized, while values close to 1 mean circularly polarization. All the waves exhibit the degree of polarization between 0.6 and 0.85, which means that they are almost circularly polarized.

\subsection{Strahl widths}
In this section we statistically compare the strahl widths for times when the whistler waves were present versus when they were absent. 

In total we found 37 time intervals during which the whistler waves were present in the B-field spectra for at least one minute and all the required wave and particle datasets are available. We also selected 31 time intervals that were adjacent to or at least very close to the first 37, but during which the whistler waves were not observed. This is a control sample. The reason that these intervals are fewer is because on some days the Cluster entered the pristine solar wind on three occasions and observed the whistlers on two of them. The remaining time interval was then used for a control sample. Both samples are required in order to compare the properties, such as the solar wind ion moments, the IMF strength, the strahl widths and the electron moments  during times when the whistlers are present and when they are not. All the intervals are listed in Table~\ref{tab1}.

The number of selected whistler intervals is not very large considering the long time period during which they were found. There are several reasons for this. The Cluster spacecraft do not spend much time in the pristine solar wind, we eliminated all the intervals when structures, such as interplanetary coronal mass ejections (ICME) and stream interaction regions (SIR) were present in the solar wind and also several different datasets (electron and ion data, magnetic field measurements and STAFF and PEACE datasets) all had to be available for the purpose of this study.
For comparison, \citet{lacombe14} report the presence of whistler waves in $\sim$10~\% of their selected data. However these authors did not check for the presence of ICMEs or SIRs in the solar wind. \citet{breneman10} for example found that intense whistler waves are most commonly found within the SIRs, close to IP shocks and near the heliospheric current sheet crossings.

We first divide each interval from Table~\ref{tab1} into consecutive one minute subintervals and calculate the average PADs during those times. Each PAD obtained this way is considered as one measurement in our samnple. By doing this we give more weight to longer time intervals and less to shorter ones.
Next we fit these average distributions with the function explained in Equation~\ref{eqn:02}. We do this for 
twelve central energies for times with and without the whistlers and compare them.
It should be noted that the total number of samples is different for different energies. In some cases the data was too noisy to allow the fitting. We visually inspect each fit in order to approve or reject it. The actual number of samples at each central energy is shown in Table~\ref{tab2}. Finally we calculate the average values and the error of the mean of the strahl width at each electron energy.

\begin{figure*}
\includegraphics[width = 0.7\textwidth]{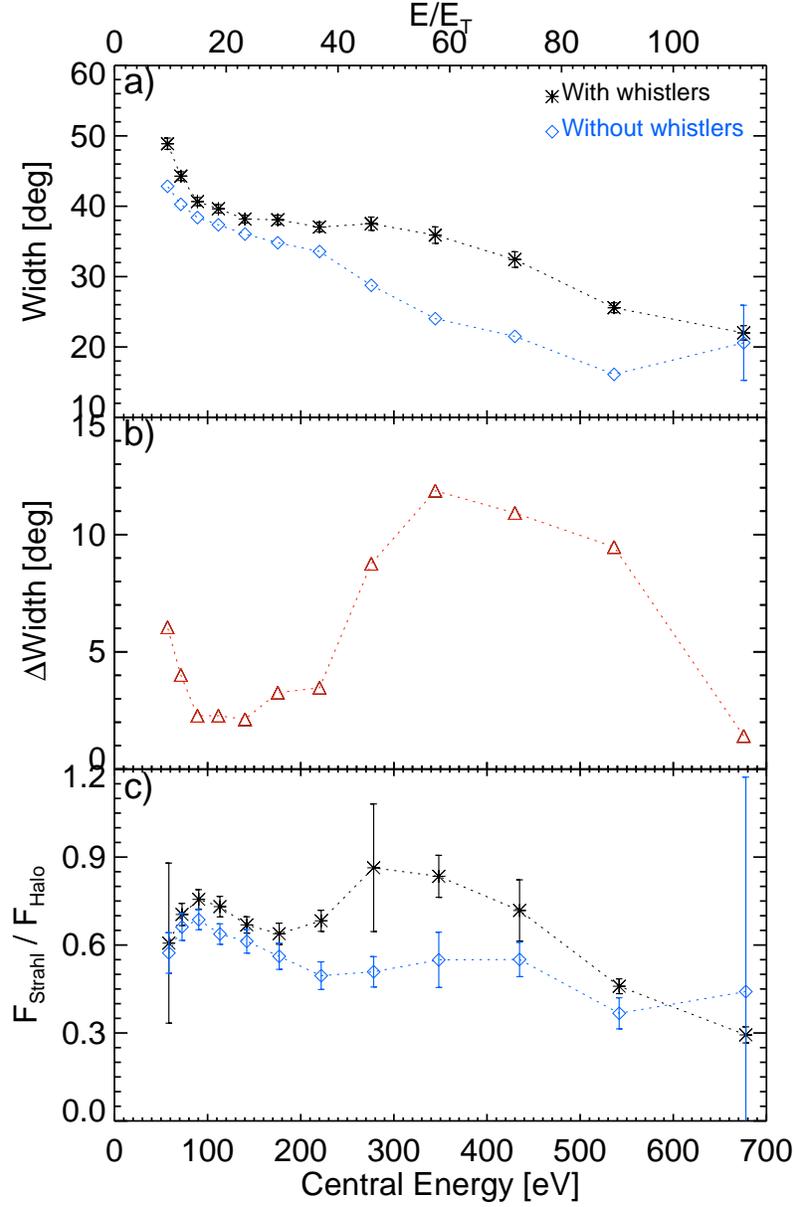}
\caption{a) Average strahl widths at times with the whistler waves (black asterisks) and during times when they were absent (blue diamonds). b) Difference in strahl widths. c) Strahl to halo flux ratios at times with and without the whistler waves.}
\label{fig:results}
\end{figure*}

Figure~\ref{fig:results} exhibits the results of this comparison. On panel a) we show the average strahl widths in presence (black asterisks) and absence (blue diamonds) of the whistler waves. We also plot the error bars indicating the error of the mean of each sample (the spread of the distribution). While on the lower $x$-axis we show electron energy in units of eV, on the upper abscissa we show electron energy normalized to typical electron thermal energy E$_T$ in the solar wind. In order to calculate the latter we assumed a typical electron temperature in the solar wind to be 140,000~K, which corresponds to E$_T$=$\frac{1}{2}k_BT$$\sim$6~eV (k$_B$ is the Boltzmann constant). This electron temperature has been shown to be a very good approximation \citep{newbury96} independent of other solar wind parameters including the proton temperature \citep{feldman77, newbury95}.

What can be seen in Figure~\ref{fig:results}a is that in the absence of the whistler waves {\bf (blue diamonds)}, the strahl width diminishes monotonically with increasing energy. The only exception is the width at 676~eV ($\sim$113~E$_T$), for which the number of one minute subintervals is smallest and the errors of the mean are largest. In the presence of the whistlers {\bf (black asterisks)} the strahl width remains roughly constant between $\sim$111~eV ($\sim$19~E$_T$) and $\sim$344~eV ($\sim$57~E$_T$). We should emphasize that this behaviour is only observed on average. The behaviour of the strahl width varies from case to case (not shown) and can be roughly constant or can diminish with increasing energy.
In the past there have been some works that have reported different variations  of the strahl width. \citet{pilipp87a, pilipp87b}, \citet{feldman78, feldman82}, \citet{hammond96} and \citet{fitzenreiter98} reported diminishing widths of the strahl as a function of energy, similar to our case without whistler mode waves. On the other hand, the strahl width has been found to increase with energy in the presence of enhanced magnetic fluctuations, possibly whistler mode waves at frequencies $\lesssim$3~Hz \citep{pagel07}. \citet{anderson12} found that broadening or widening of the strahl with energy occur with equal probability. 

Figure~\ref{fig:results}a also shows that the average strahl widths in the presence of the whistlers are larger than when the whistlers are absent. This is true at all energies. The difference in the average strahl widths ($\Delta w$) varies strongly with the electron energy (Figure~\ref{fig:results}b) {\bf but is always larger than the error bars, except at 676~eV}. At 57~eV ($\sim$10~E$_T$) the difference in PA is 6$^\circ$, it diminishes to 2$^\circ$ between 89~eV ($\sim$15~E$_T$) and 140~eV ($\sim$23~E$_T$) and then it starts rising again. It reaches the maximum of 12$^\circ$ at 344~eV.

Next we show that the largest strahl widths occur preferentially when the differential energy flux (DEF) integrated over all pitch angles of the strahl is more intense relative to the halo DEF. Figure~\ref{fig:results}c shows how the ratios of the strahl and the halo DEFs (F$_{strahl}$/F$_{halo}$) vary with energy in cases when the whistlers were present and when they were absent. {\bf In each individual case this ratio depends on central energies and energy ranges of the strahl and the halo components, so only average values of a larger sample of events can tell us whether this ratio is different when the whistler mode waves are present. We see in Figure~\ref{fig:results}c that regardless of the presence of whislers the average F$_{strahl}$/F$_{halo}$ ratio incrases for energies between 57~eV and 89~eV and then it decreases for energies up to 175~eV ($\sim$29~E$_T$).}
After that the F$_{strahl}$/F$_{halo}$ ratio for times when there were no whistler waves remains roughly constant.
During times when the whistlers were present however this ratio increases, reaching a peak at $\sim$280~eV ($\sim$47~E$_T$) and it decreases afterwards. The difference of F$_{strahl}$/F$_{halo}$ for times with and without the whistlers tends to be larger at energies at which the $\Delta w$ is also larger (compare with Figure~\ref{fig:results}b).

{\bf The error bars in Figure~\ref{fig:results}c are generally small and do not overlap. This means that the difference in F$_{strahl}$/F$_{halo}$ at most energies (with exceptions at 57~eV and 676~eV) is larger than the measurement errors.}

\section{Discussion}
\label{sec:dis}
In this work we perform a statistical study of the widths of suprathermal strahl in pristine solar wind  at 1~AU in the presence of narrow-band quasi-parallel ($k_{\|} \gg k_{\perp}$) whistler waves observed within $[5,100]$~Hz frequency range. 
The strahl widths during the time intervals with such whistlers waves are compared to the strahl widths during times when the whistlers are absent. To our knowledge this is the first observational study that is trying to relate the narrow-band whistler waves in the pristine solar wind to the widening of the suprathermal electron strahl component.

In the past the so called broad-band whistler waves have been invoked in order to explain the strahl broadening. \citet{pagel07} report observations of broad strahl pitch-angle distributions during times of enhanced magnetic field fluctuations at $\leq$3~Hz. During those intervals the strahl PA widths were broader than at earlier or later times. The authors used the Advanced Composition Explorer (ACE) B-field measurements with time resolution of 3 vectors per second. They concluded that strahl PA broadening was due to broad-band magnetic field whistler fluctuations, although the authors did not explicitly show that the whistler mode waves with broadband spectra and with quasi-parallel wave vectors able to interact resonantly with electrons, were actually present during the observed time intervals. Numerical simulation results by \citet{vocks05} and theoretical considerations by \citet{pierrard11} have shown if with sunward propagating whistler waves with wave vectors parallel to the background magnetic field consituted the background turbulence, then their interaction with the suprathermal electrons would result in strahl broadening.

However there is a problem when talking about the ``broad-band whistler waves'' at sub-ion scales.  Within this frequency range whistler waves may exist and very often in the literature the authors call this range a ``whistlers range''. Whether such whistlers actually exist is still an open question. Observations show that the background turbulence in the solar wind at sub-ion scales are mainly transverse with the power in the direction perpendicular to the magnetic field ($\delta B_\perp^2$) larger than the power in the parallel direction ($\delta B_\parallel^2$) by as much as a factor of 20 \citep{chen10}. Both components (perpendicular and parallel) exhibit $k_\perp > k_\parallel$ \citep{chen10}. The fact that wave vectors of the background turbulence are mainly perpendicular to the magnetic field was also shown by \citet{mangeney06} and \citet{alexandrova08}. The general behavior of the solar wind background turbulence (with $k_{\perp}\gg k_{\|}$) at sub-ion scales was studied by \citet{alexandrova12}. Although these  observations are consistent with whistler mode waves, they are also consistent with kinetic Alfv\'en waves. However even if they are whistler mode waves, in order for them to efficiently interact with parallel propagating strahl electrons, they are required to have wave vectors parallel to the background magnetic field. Hence the observations of the electromagnetic turbulence in the solar wind are not favourable to the idea of broad-band whistlers scattering the strahl electrons.

In this paper we show that narrow-band whistler mode waves can efficiently interact with the strahl electrons. These waves are observed at around $0.1 f_{ce}$ (or within [5,90]~Hz frequency range) in the solar wind and their wave vectors are almost parallel to the background B-field. Their durations had to be at least 1 minute in the spacecraft data in order to include them in our sample. The time intervals containing these waves have been chosen so the four Cluster spacecraft were in the pristine solar wind, so not inside the ICMEs, SIRs or the Earth's foreshock. We also excluded any intervals when the spacecraft were close to IP shocks or when electron distributions such as conics or shoulders were present.

We show that narrow-band whistlers modify the dependence of the strahl widths as a function of electron energy. 
The strahl broadening occurs at all energies, but its magnitude is different at different electron energies and ranges between 2$^\circ$ and 12$^\circ$ PA. Strahl widths do no longer diminish monotonically as a function of the electron energy (as is the case in the absence of the narrow-band whistlers). On average, strahl widths diminish at energies below $\sim$111~eV ($\sim$19~E$_T$), then remain roughly constant and even sightly increase for E$\lesssim$276~eV ($\sim$46~E$_T$) and then they diminish again at higher energies. The largest difference between the average strahl width in the presence and absence of the whistler mode waves occured at E$\sim$344~eV ($\sim$57~E$_T$) and is equal to $\sim$12$^\circ$ PA.

This energy dependence of the strahl width is different from what was reported in the past. \citet{pilipp87a}, \citet{pilipp87b}, \citet{feldman78}, \citet{feldman82}, \citet{hammond96} and \citet{fitzenreiter98} studied strahl widths in the solar wind as a function of electron energy and concluded that the widths diminish with increasing energy of electrons. We also see in our Figure~\ref{fig:results}a that when the narrow-band whistler mode waves are not present, the strahl narrows monotonically with increasing electron energy. \citet{pagel07} reported increasing strahl widths as a function of electron energy in the presence of low frequency ($\leq$3~Hz) magentic field fluctuations.
However, as we show here whistlers with parallel wave vectors are observed at $f>3Hz$. Therefore the role of whistlers in the observations of  \citet{pagel07} is questionable.

The F$_{strahl}$/F$_{halo}$ ratio (Figure~\ref{fig:results}c) at times without the whistlers shows very little dependence on the electron energy. At times when the whistlers were present this ratio is increased for energies between $\sim$220~eV ($\sim$37~E$_T$) and $\sim$536~eV ($\sim$89~E$_T$) and peaks at $\sim$280~eV ($\sim$47~E$_T$). In order to interact with strahl electrons in this energy range the whistler phase velocities need to be between $\sim$970~kms$^{-1}$ and $\sim$1500~kms$^{-1}$ (taking the whistler frequency $\omega$=100~Hz, $\Omega_e$=1000~Hz and parallel propagation of the strahl electrons). The difference of F$_{strahl}$/F$_{halo}$ ratios for times with and without the whistlers tends to be larger at energies at which the strahl width is also larger. This suggests that the more intense strahl relative to the core is related to wider strahls and to the presence of the whistler mode waves. 

\section{Conclusions}
In this work we show that narrow-band whistler mode waves do interact with the strahl electrons. This interaction results in the broadening of the stragl PA width, which is different at different energies. The largest strahl broadning occured at electron energy  of E$\sim$344~eV ($\sim$57~E$_T$) and was equal to $\sim$12$^\circ$ PA. The dependence of strahl width as a function of energy is modified in the presence of the whistlers since the width no longer decreases monotonically with increasing energy as is observed in the absence of quasi-parallel propagating whistler waves. During times when the narrow-band whistlers are present, the ratio of strahl to halo fluxes F$_{strahl}$/F$_{halo}$ is also increased. The more intense strahl relative to the core is related to larger strahl widths and to the presence of the whistler mode waves. 

The question arises how much the whistler mode waves in the interplanetary space contribute to strahl scattering in order to account for the formation of the halo component. Our study is performed with the Cluster data at 1~AU and it does not show the accumulative effect that such interactions could have along the entire electron's trajectory from Sun to 1~AU. However the broadening of strahl by up to 12$^\circ$ PA suggests that it is plausible that narrow-band whistler waves contribute importantly to the overall broadening of the suprathermal electron strahl.

\acknowledgments
The authors are grateful to the Cluster Science Archive teams (ESA) and the CL/CLWeb team (IRAP) for the easy access to the Cluster data; to O. Santolik for the PRASSADCO program which 
gives the polarization and propagation properties of the magnetic fluctuations  measured by STAFF-SA and to P. Robert for a polarization program of waveform data, STAFF-SC, and FGM.
The authors also acknowledge the Cluster teams: STAFF (PIs P. Canu and N. Cornilleau-Wehrlin), FGM (PIs C. Carr, A. Balogh and E. Lucek); CIS (PIs H. R\`eme and I. Dandouras); PEACE (PI A. Fazakerley). The CLUSTER STAFF instrument has been developed and operated with the support of CNES and CNRS. PK's work was supported by the PAPIIT grant IA104416.
 
\email{aastex-help@aas.org}.

{\it Facilities:} \facility{CSA}.

\clearpage

\clearpage


\clearpage

\clearpage

\clearpage

\begin{deluxetable}{cccccc}
\tabletypesize{\scriptsize}
\tablecaption{Sampled time intervals}
\tablewidth{0pt}
\tablehead{\colhead{With whistlers} & & & \colhead{Without whistlers} & & \\
\colhead{Date} & \colhead{Time UT} & \colhead{Spacecraft} &\colhead{Date} & \colhead{Time UT} & \colhead{Spacecraft}\\
\colhead{YYYY/MM/DD} & \colhead{hr:mn} & STAFF, FGM, DEF/Ion moments &\colhead{YYYY/MM/DD} & \colhead{hr:mn} & STAFF, IMF, DEF/Ion moments}
\startdata
2001/02/19 & 17:15-17:19 & C1/C1 & 2001/02/19 & 19:22-19:29 & C1/C1 \\
2001/02/19 & 17:21-17:42  & C1/C1 & 2001/02/19 & 19:55-20:10 & C1/C1 \\
2002/02/09 & 02:30-02:40  & C1/C1 & 2002/02/09 & 02:10-02:20 & C1/C1 \\
2003/01/30 & 00:38-01:35  & C1/C1 & 2003/01/30 & 00:00-00:30 & C1/C1 \\
2003/01/30 & 02:42-03:10  & C1/C1 & 2003/01/30 & 03:25-03:35 & C1/C1 \\
2003/01/30 & 07:05-07:33  & C1/C1 & 2003/01/30 & 08:16-08:18 & C1/C1 \\
2003/01/30 & 07:38-07:45  & C1/C1 & 2003/01/30 & 09:15-09:25 & C1/C1 \\
2004/02/09 & 19:19-19:23  & C4/C1 & 2004/02/09 & 19:12-19:17 & C4/C1 \\
2004/02/09 & 20:36-20:43  & C4/C1 & 2004/02/09 & 19:50-20:00 & C4/C1 \\
2004/02/09 & 20:59-21:02  & C4/C1 & 2004/02/09 & 20:20-20:30 & C4/C1 \\
2004/02/09 & 22:04-22:08  & C4/C1 & 2004/02/09 & 21:50-22:00 & C4/C1 \\
2004/02/09 & 22:44-22:45  & C4/C1 & 2004/02/09 & 22:50-23:00 & C4/C1 \\
2004/04/18 & 11:40-12:15  & C1/C1 & 2005/02/16 & 11:50-12:00 & C1/C1 \\
2005/02/16 & 11:19-11:22  & C1/C1 & 2007/03/04 & 07:00-07:10 & C1/C1 \\
2007/03/04 & 07:18-07:26  & C1/C1 & 2009/02/08 & 04:00-04:30 & C2/C1 \\
2009/02/08 & 04:33-06:18  & C2/C1 & 2009/02/20 & 03:00-03:10 & C2/C1 \\
2009/02/20 & 03:38-03:48  & C2/C1 & 2009/02/20 & 04:50-05:00 & C2/C1 \\
2009/02/20 & 03:48-04:05  & C2/C1 & 2009/02/21 & 14:50-15:00 & C2/C1 \\
2009/02/21 & 14:10-14:39  & C2/C1 & 2009/02/21 & 15:45-15:55 & C2/C1 \\
2009/02/21 & 15:20-15:27  & C2/C1 & 2009/04/27 & 04:40-04:50 & C2/C1 \\
2009/04/27 & 04:53-05:21  & C2/C1 & 2010/01/11 & 20:00-20:10 & C2/C1 \\
2010/01/11 & 20:17-20:21  & C2/C1 & 2010/01/11 & 20:30-20:38 & C2/C1 \\
2010/01/11 & 20:25-20:27 & C2/C1 & 2010/02/23 & 17:13-17:20 & C2/C1 \\
2010/02/23 & 17:03-17:13 & C2/C1 & 2010/02/25 & 14:25-14:35 & C2/C1 \\
2010/02/25 & 13:37-13:57 & C2/C1 & 2010/02/25 & 20:58-21:03 & C2/C1 \\
2010/02/25 & 20:10-20:58 & C2/C1 & 2010/02/28 & 11:15-11:25 & C2/C1 \\
2010/02/25 & 21:09-21:47 & C2/C1 & 2010/03/15 & 22:40-22:50 & C2/C1 \\
2010/02/28 & 11:04-11:09 & C2/C1 & 2010/03/15 & 23:18-23:20 & C2/C1 \\
2010/03/15 & 23:03-23:05 & C2/C1 & 2010/03/16 & 01:35-01:39 & C2/C1 \\
2010/03/15 & 23:26-23:37 & C2/C1 & 2010/04/18 & 20:01-20:07 & C2/C1 \\
2010/03/15 & 23:40-23:55 & C2/C1 & 2010/04/19 & 04:45-04:55 & C2/C1 \\
2010/03/15 & 23:58-24:00 & C2/C1 & & & \\  
2010/03/16 & 00:00-00:23 & C2/C1 & & & \\  
2010/03/16 & 00:29-00:38 & C2/C1 & & & \\  
2010/03/16 & 01:29-01:34 & C2/C1 & & & \\  
2010/04/18 & 19:40-19:54 & C2/C1 & & & \\  
2010/04/19 & 04:10-04:18 & C2/C1 & & &   
\enddata
\label{tab1}
\end{deluxetable}

\clearpage

\begin{deluxetable}{ccc}
\tabletypesize{\scriptsize}
\tablecaption{Number of samples of suprathermal electrons at different energies}
\tablewidth{0pt}
\tablehead{\colhead{Central energy (eV)} & \colhead{No. of samples with whistlers} & \colhead{No. of samples without whistlers}}
\startdata
676 & 48 & 22\\
536 & 88 & 30\\
430 & 114 & 38\\
344 & 136 & 43\\
276 & 163 & 66\\
220 & 217 & 101\\
175 & 275 & 123\\
140 & 350 & 156\\
111 & 398 & 197\\
89 & 441 & 220\\
71 & 469 & 233\\
57 & 423 & 213\\
\enddata
\label{tab2}
\end{deluxetable}

\end{document}